# Analysis and Modeling of Microstructured Fiber Using The Analytical Method Based on The Empirical Equation

DEBBAL Mohammed[1], CHIKH-BLED Mohammed[2]

[1] University of Tlemcen, Algeria,
Department of electrical Engineering, Faculty of Technology, Telecommunication Laboratory.
E-Mail1: debbal.mohammed@gmail.com

[2] University of Tlemcen, Algeria,
Department of electrical Engineering, Faculty of Technology, Telecommunication Laboratory.

*Abstract* – **In this paper, a new class of optical fibers is studied, i.e. microstructured fibers or photonic crystal fibers (PCF). The main objective is to characterize these fibers using different dispersion diagrams and present an interface that enables calculation of various transverse indexes. The choice of these profiles was guided by a simplification achievement. Furthermore, in order to acquire expertise in air-silica microstructured fibers, a behavior modeling of these particular structures is proposed in this paper in order to present an analytical method based on empirical equations. The fibers treated in this paper have a triangular arrangement of holes corresponding to the natural arrangement of a bundle of cylindrical tubes around a central rod of the same diameter.**

## I. INTRODUCTION

The idea of photonic crystal fibers (PCF) dates back to 1991, but the production of the first PCF dates only from 1995. Problems related to the production of such fibers were committed to the University of Bath by the company's founders Blaze Photonics.

PCF fibers are no different from conventional fibers at first. However, the microscopic study of the section of a PCF can observe a particular structure. Indeed, the PCF fibers are constructed from a hexagonal structure of small cylinders of air around a core that can be made of silica or air, depending on the use you intend to make.

Air hole in a regular triangular lattice with diameter (d), the space between these air holes is the pitch Λ (center-to-center spacing)

Photonic Crystal Fibers (PCFs) [1]–[2], have been under intensive study for the past several years as they offer a number of unique and useful properties not achievable in standard silica glass fibers. PCFs fall into two basic categories:

The first one, an index-guiding PCF [3], [4], is usually formed by a central solid defect region surrounded by multiple air holes in a regular triangular lattice and confines light by total internal reflection like standard fibers.

The second one uses a perfect periodic structure exhibiting a photonic band-gap (PBG) effect at the operating wavelength to guide light in a low index core region, which is also called PBG fiber (PBGF) [5], [6].

In the next part we are going to see the different propagation index, by analyzing and modeling this index throw the empirical equation, and by the end of this paper we present our interface that we made it to help us to calculate the different propagation index.

## II. THE NORMALIZED FREQUENCY

The normalized frequency 'V' is a parameter that contributes to characterize the guiding conditions in standard fibers.

According to Gloge formula (Eq.1) by plotting the normalized propagation constant 'b' [7] of each of the modes propagating in the fiber as a function of V, we obtains the normalized dispersion curve for each of these modes.

$$b = \frac{n_{eff}^2 - n_{gaine}^2}{n_{coeur}^2 - n_{gaine}^2} = 1 - \left[\frac{1+\sqrt{2}}{1+(4+V^4)^{1/4}}\right]^2 \quad (1)$$

The normalized frequency depends on the wavelength in vacuum 'λ0', fiber radius core 'a', and index core '$n_c$' and cladding index '$n_g$'. The normalized frequency is expressed by (Eq.2):

$$V = \frac{2\pi}{\lambda_0} a \sqrt{n_c^2 - n_g^2} \quad (2)$$

The cutoff normalized frequency associated with the cutoff wavelength of a mode is that for which $n_{eff} = n_{gaine}$ (b=0). The cutoff wavelength 'λc' it the wavelength below which the fiber is not single-mode.

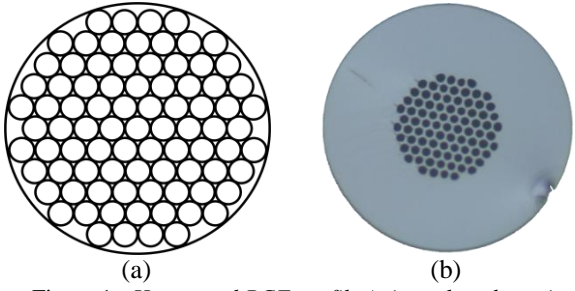

Figure 1: *Hexagonal PCF profile (triangular photonic cladding).*
*(a) Transverse arrangement of capillaries PCF preform.*
*(b) Transverse index profile of the fiber obtained from the perform (white: silica index = n (λ), black: air index = 1).*

As in the standard fiber, the spectral propagation in a monomode PCF is the whole of wavelengths for which V is below the cutoff normalized frequency of the second mode.

For PCF, in most of the work that can be found in the literature, it appears that the description of the optical properties of photonic crystal fibers (PCF) is mainly based on numerical approach. Birks and al were the first to demonstrate that the microstructured fiber has a behavior similar to that of a step index fiber (FSI) [8]. In PCF fibers, the normalized frequency V is defined by (Eq.3):

$$V_{PCF} = \frac{2\pi}{\lambda} a_{eff} \sqrt{n_{co}^2 - n_{FSM}^2} \quad (3)$$

Where 'λ' it the wavelength in vacuum, '$a_{eff}$' is the equivalent core radius of the PCF and '$n_{co}$' it the refractive core index, '$n_{FSM}$' is the refractive cladding index.
V will determine if the fiber is multimode or singlemode.
✓ If 'V < 2.405', one mode propagates in the fiber, the mode 'HE11' or 'LP01', also known as the fundamental mode of the fiber. The fiber is called single-mode.
✓ If 'V > 2.405', several modes can propagate, the fiber is called multimode.

### III. THE RADIUS CORE OF THE PCF

In the context of photonic crystal fibers, it is natural to consider the 'V' parameter, which was done in the seminal work by the group from the University of Bath [3] and in subsequent work on the singlemode properties [8] and the actual values of the effective frequency 'Veff' [10]. However, in trying to adopt equation (Eq.2) in PCF, the difficulty lies in the determination of an equivalent core radius 'a' (Eq.3).
To adapt this concept of the value of 'V' to the PCF, several effective core radius have been proposed, such as $a_{eff} = \Lambda$ [3], [8], $\Lambda/2$ [11], $2\Lambda - d$ [12], $\Lambda/\sqrt{3}$ [13] and in the references [2] and [4], it was mention that we can choose any transverse dimension.

To determinate '$a_{eff}$', we use the dispersion relation of the fundamental mode LP 01 (corresponding to the electromagnetic mode HE 11) in a fiber [14]:

$$U\frac{J_1(U)}{J_0(U)} - W\frac{K_1(W)}{K_0(W)} = 0 \quad (4)$$

In this relationship, $J_i$ and $K_i$ are the Bessel functions of the first and second kind with the order 'i'. Variables 'U' and 'W' are:

$$U = \frac{2\pi}{\lambda} a \sqrt{n_{co}^2 - n_{eff}^2} = a\sqrt{k_0^2 n_{co}^2 - \beta^2} \quad (5)$$

$$W = \frac{2\pi}{\lambda} a \sqrt{n_{eff}^2 - n_{FSM}^2} \quad (6)$$

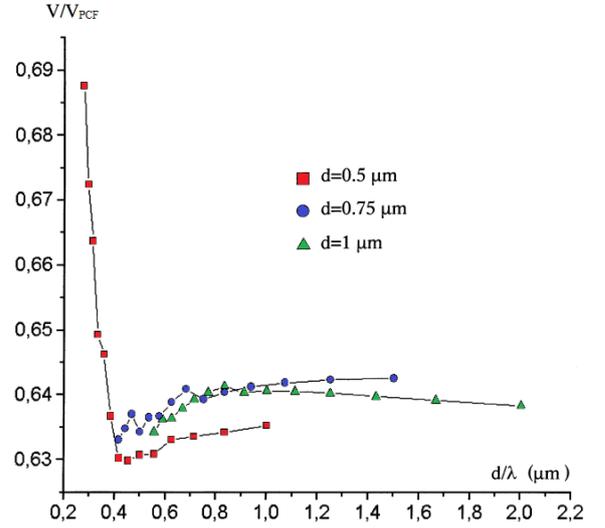

Figure 2: *Comparison of normalized frequencies V and VPCF* [15]

With '$n_{eff}$' is the effective index of the fundamental mode guided.
The parameters 'U' and 'W' are respectively called the normalized transverse propagation constants in the core (Eq.5) and cladding (Eq.6).
The value of the core radius thus appears in the dispersion relations above, but it can be eliminated by expressing 'U' and 'W' based solely on 'V' and 'b' (Eq.8) through the formula:

$$U^2 + W^2 = V^2 \Rightarrow V = \sqrt{U^2 + W^2} \quad (7)$$

We then obtain:

$$U = V\sqrt{1-b} \text{ et } W = V\sqrt{b} \quad (8)$$

In Figure 2, we see that '$V/V_{PCF}$' = '$a_{eff}/\Lambda$' is close to 0.64 for the three considered profiles and the wavelengths such as $d/\lambda > 0.45$ [15]. The choice of a core radius '$a_{eff}$' equal to 0.64 Λ is suitable when the working wavelength is less than d / 0.45.

To summaries, at a given wavelength λ, we found a step-index fiber ($n_{cœur} = n_{silice}(λ)$, $n_{cladding} = n_{eff\ cladding}(λ)$) et $a = 0,64Λ$) equivalent to the PCF considered and the value of V corresponds to a normalized propagation constant 'b' equal to the fundamental mode of the PCF. It may be noted that with the same value of V, this correspondence is maintained for higher order modes [15]. several formulations exist for the 'V' parameter, we chose the one proposed by Koshiba and al [16] (Eq. 3). In this definition. The effective core radius of the FCP is chosen so that the fiber is single mode for 'V <2.405'. This description is valid for the limit values of Λ / λ greater than 0.5 [17].

## IV. THE V PARAMETER

Although the V parameter provides a simple way to modeling the PCF [9], a factor limiting the use of the equation (Eq. 3), a numerical method is still necessary for the effective cladding index '$n_{FSM}$'. It is thus possible to apply directly the model of FSI to the microstructured fibers [18]. It would be useful to have another expression depends only on the wavelength λ, and the parameters of the geometric structure 'd' and 'Λ'.

Nielsen and Mortensen are proved as [9] that V parameter can approximate with a function of type Eq. 5 depending on the wavelength λ, and the structural parameters d and Λ.

$$V\left(\frac{\lambda}{\Lambda}, \frac{d}{\Lambda}\right) = \frac{A\left(\frac{d}{\Lambda}\right)}{B\left(\frac{d}{\Lambda}\right) \times \exp\left[C\left(\frac{d}{\Lambda}\right) \times \frac{\lambda}{\Lambda}\right] + 1} \quad (9)$$

In figure 3 we show V as function of λ/Λ for d/Λ ranging from 0.20 to 0.80 in steps of 0.1.
The parameter A, B and C are depend on d/λ only, and are describing by the following expressions:

$$A\left(\frac{d}{\Lambda}\right) = \frac{d}{\Lambda} + 0.457 + \frac{3.405 \times \frac{d}{\Lambda}}{0.904 - \frac{d}{\Lambda}} \quad (9.1)$$

$$B\left(\frac{d}{\Lambda}\right) = 0.200 \times \frac{d}{\Lambda} + 0.100 + 0.027 \times \left(1.045 - \frac{d}{\Lambda}\right)^{-2.8} \quad (9.2)$$

$$C\left(\frac{d}{\Lambda}\right) = 0.630 \times \exp\left(\frac{0.755}{0.171 + \frac{d}{\Lambda}}\right) \quad (9.3)$$

Equation (Eq.9) is the empirical expression for the 'V' parameter in a PCF with 'λ/Λ' and 'd/Λ' is the only input parameter (Figure 3). For (λ/Λ <2) and (V> 0.5) the expression gives a values of 'V' which deviates less than 3% of the correct values obtained from equation (Eq.3) [9].

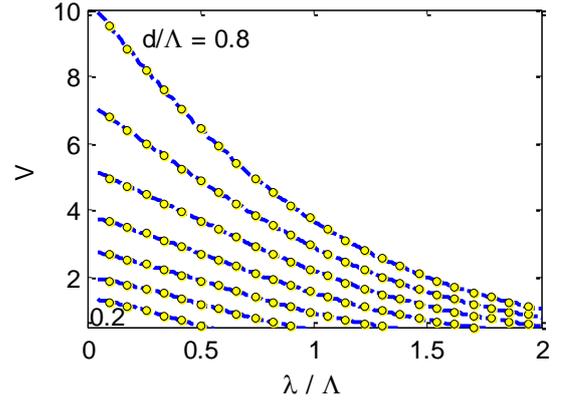

Figure 3: *V as a function of relative wavelength λ/Λ for d/Λ ranging from 0.20 to 0.80 in steps of 0.1*

Saitoh and al have shown [17] that 'V' can be approximated by another function of type (Eq.10) according to the wavelength λ, and the structure parameters 'd' and 'Λ':

$$V\left(\frac{\lambda}{\Lambda}, \frac{d}{\Lambda}\right) = A_1 + \frac{A_2}{1 + A_3 \exp\left(A_4 \frac{\lambda}{\Lambda}\right)} \quad (10)$$

with

$$A_i = a_{i0} + a_{i1}\left(\frac{d}{\Lambda}\right)^{b_{i1}} + a_{i2}\left(\frac{d}{\Lambda}\right)^{b_{i2}} + a_{i3}\left(\frac{d}{\Lambda}\right)^{b_{i3}} \quad (10.1)$$

Table 1 : *Coefficients used in [19] to calculate the normalized frequency by the analytical method*

|          | i = 1     | i = 2     | i = 3    | i = 4    |
|----------|-----------|-----------|----------|----------|
| $a_{i0}$ | 0.54808   | 0.71041   | 0.16904  | -1.52736 |
| $a_{i1}$ | 5.00401   | 9.73491   | 1.85765  | 1.06745  |
| $a_{i2}$ | -10.43248 | 47.41496  | 18.96849 | 1.93229  |
| $a_{i3}$ | 8.22992   | -437.50962| -42.4318 | 3.89     |
| $b_{i1}$ | 5         | 1.8       | 1.7      | -0.84    |
| $b_{i2}$ | 7         | 7.32      | 10       | 1.02     |
| $b_{i3}$ | 9         | 22.8      | 14       | 13.4     |

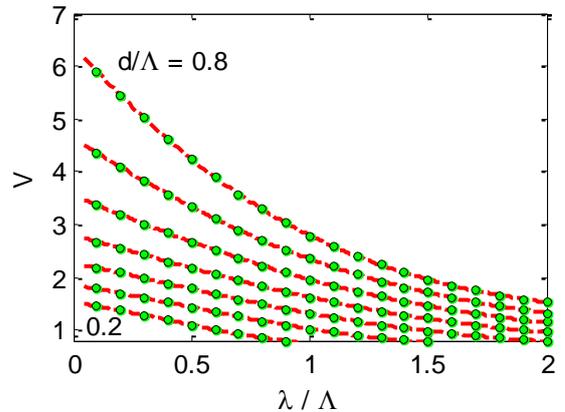

Figure 4: *V depending on the relative length λ/Λ for d/Λ range from 0.20 to 0.80 in steps of 0.1*

And where the coefficients '$a_{i0}$' to '$a_{i3}$' and '$b_{i1}$' to '$b_{i3}$' are given in the table1.

Equation (Eq.10) is the empirical expression for the 'V' parameter in a PCF with '$\lambda/\Lambda$' and '$d/\Lambda$' is the only input parameter (Figure 4). For ($\lambda/\Lambda$ <2) and (V> 0.80) the expression gives a values of 'V' which deviates less than 1.3% of the correct values obtained from equation (Eq.3) [17].

## V. THE $N_{FSM}$ INDEX

Once the wavelength λ are fixed, and the geometrical parameters 'Λ' and 'd' are known, the combination of equations (Eq. 3) and (Eq. 10) allows us to calculate the effective cladding index '$n_{FSM}$', without the use of methods and numerical simulations.

The figure 5 shows the effective cladding index '$n_{FSM}$', depending on the relative length, $\lambda/\Lambda$ for $d/\Lambda$ range from 0.20 to 0.80 in steps of 0.1 with '$a_{eff} = \Lambda/\sqrt{3}$'.

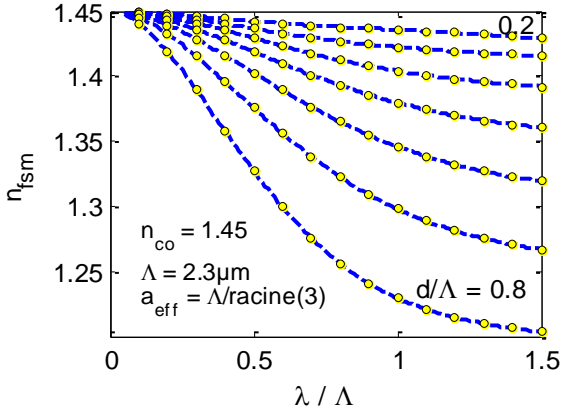

Figure 5: *The effective cladding index '$n_{FSM}$' according to the relative length λ/Λ for d/Λ range from 0.20 to 0.80 in steps of 0.1*

## VI. THE W PARAMETER

In the previous section, we have provided the empirical setting for the 'V' parameter of PCF. Using equation (Eq.10) we can easily obtain the effective cladding index '$n_{FSM}$', however, we usually need heavy numerical calculations to obtain the exact values of the effective index of the fundamental mode '$n_{eff}$' in the equation. (Eq.6).

It would be more convenient to have an empirical relation for the W parameter of the PCF.

Nielsen and al reported the empirical relation for the W parameter [19] However, we cannot get the value of '$n_{eff}$' from only the W parameter. To obtain '$n_{eff}$' we need empirical relation for the two parameters V (Eq.10) and W (Eq.11).

$$W\left(\frac{\lambda}{\Lambda},\frac{d}{\Lambda}\right) = B_1 + \frac{B_2}{1 + B_3 \exp\left(B_4 \frac{\lambda}{\Lambda}\right)} \qquad (11)$$

With

$$B_i = c_{i0} + c_{i1}\left(\frac{d}{\Lambda}\right)^{d_{i1}} + c_{i2}\left(\frac{d}{\Lambda}\right)^{d_{i2}} + c_{i3}\left(\frac{d}{\Lambda}\right)^{d_{i3}} \qquad (11.1)$$

And where the coefficients '$c_{i0}$' to '$c_{i3}$' and '$d_{i1}$' to '$d_{i3}$' are given in the table 2.

Figure 6 shows the W parameter according to the relative length, $\lambda/\Lambda$ for $d/\Lambda$ range from 0.20 to 0.80 in steps of 0.1.

Table 2 : *Coefficients of the empirical equation used in [19] to calculate the W parameter by the analytical method*

|          | i = 1      | i = 2       | i = 3     | i = 4    |
|----------|-----------|-------------|-----------|----------|
| $c_{i0}$ | -0.0973   | 0.53193     | 0.24876   | 5.29801  |
| $c_{i1}$ | -16.70566 | 6.70858     | 2.72423   | 0.05142  |
| $c_{i2}$ | 67.13845  | 52.04855    | 13.28649  | -5.18302 |
| $c_{i3}$ | -50.25518 | -540.66947  | -36.80372 | 2.7641   |
| $d_{i1}$ | 7         | 1.49        | 3.85      | -2       |
| $d_{i2}$ | 9         | 6.58        | 10        | 0.41     |
| $d_{i3}$ | 10        | 24.8        | 15        | 6        |

The joint use of equations (Eq.10) giving the 'V' parameter and (Eq.11) giving the 'W' parameter, allow to calculate the effective index of the fundamental mode '$n_{eff}$' without the need of numerical simulations.

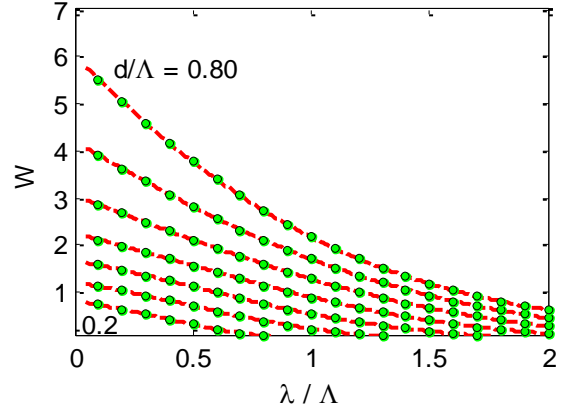

Figure 6: *W parameter depending on the relative length λ/Λ for d/Λ ranging from 0.20 to 0.80 in steps of 0.1.*

## VII. THE $N_{eff}$ INDEX

In the previous section we deduce from equations (Eq.3) and (Eq.10) the effective cladding index '$n_{FSM}$'. Finally, '$n_{FSM}$' is known, and from equations (Eq.6) and (Eq.11) we obtain the effective index of the fundamental mode '$n_{eff}$'.

The figure 7 shows the effective index of the fundamental mode '$n_{eff}$', according to the relative length $\lambda/\Lambda$ for $d/\Lambda$ ranging from 0.20 to 0.80 in steps of 0.1 with '$a_{eff} = \Lambda/\sqrt{3}$'.

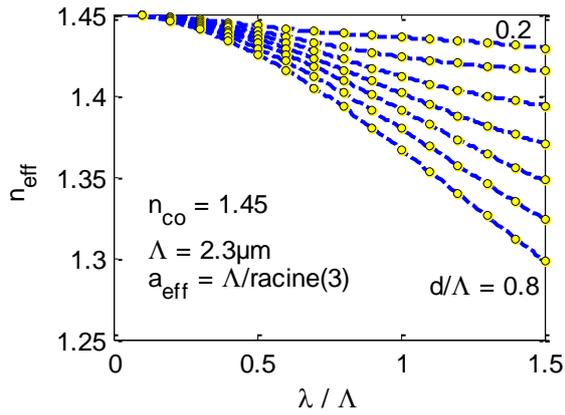

Figure 7: *The effective index of the fundamental mode '$n_{eff}$', depending on the relative length λ/Λ for d/Λ ranging from 0.20 to 0.80 in steps of 0.1*

## VIII. DESIGN OF AN INTERFACE FOR CALCULATION

In order to modeling the photonic crystal fiber using the analytical method, based on empirical equations that we discussed in the previous section, we have designed an interface that allows us to facilitate the calculations and plot the following parameters:
- The V parameter of the fiber ($V_{eff}$).
- The W parameter of the fiber ($W_{eff}$).
- The effective cladding index ($n_{PCF}$).
- The effective index of the fundamental mode ($n_{eff}$).

This interface allows us to modeling the fiber as a function of different values of effective core radius ($a_{eff}$), with the ability to export the graph in image format (Figure 8 and 9).

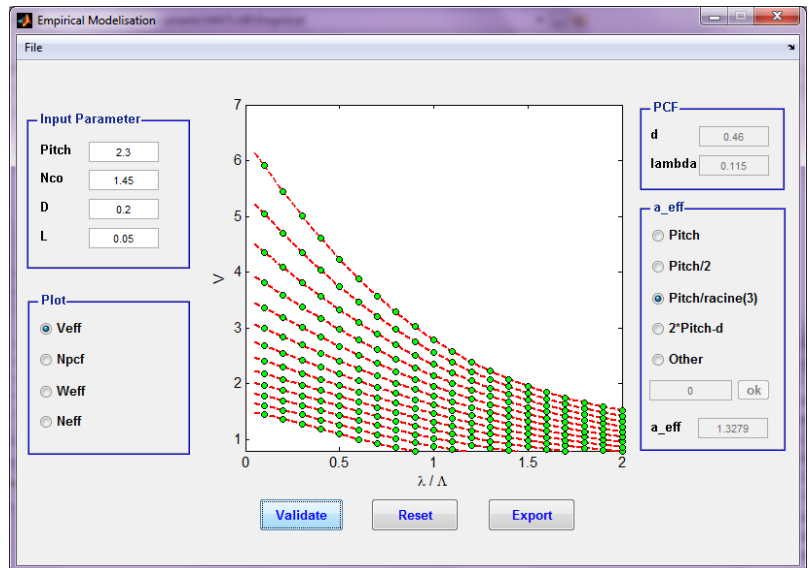

Figure 8: *V parameter calculated by the interface $\left(a_{eff} = \frac{\Lambda}{\sqrt{3}}\right)$*

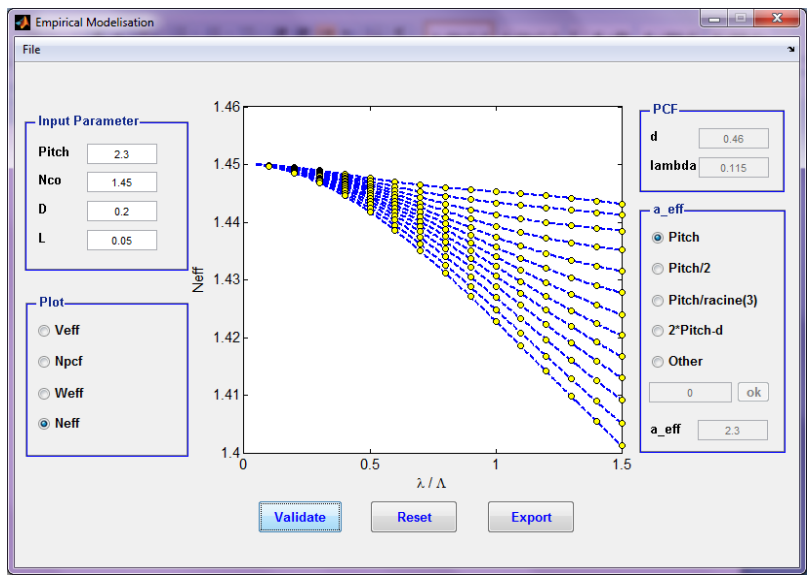

Figure 9: *Neff index calculated by the interface ($a_{eff} = \Lambda$)*

## IX. CONCLUSIONS

Photonic crystal fibers combine properties of 2D photonic crystals and classical fibers. Research on photonic crystal fibers is still very young and we may expect many new developments, more accurate and efficient methods for designing and optimization.

We have established an analytical study of optical properties of microstructured fibers in order to modeling and characterize the properties of various propagation indices, and made an interface that allows us to calculate these indices.